\documentclass[prb,twocolumn,showpacs,floatfix]{revtex4}
\usepackage{graphicx}

\begin{document}

\title{
   Three-body correlations and finite-size effects
   in the Moore--Read states on a sphere}

\author{
   Arkadiusz W\'ojs$^{1,2}$ and 
   John J. Quinn$^1$}

\affiliation{
   $^1$University of Tennessee, Knoxville, Tennessee 37996, USA\\
   $^2$Wroclaw University of Technology, 50-370 Wroclaw, Poland}

\begin{abstract}
Two- and three-body correlations in partially filled degenerate 
fermion shells are studied numerically for various interactions 
between the particles.
Three distinct correlation regimes are defined, depending on the 
short-range behavior of the pair pseudopotential.
For pseudopotentials similar to those of electrons in the first 
excited Landau level, correlations at half-filling have a simple 
three-body form consisting of the maximum avoidance of the triplet 
state with the smallest relative angular momentum $\mathcal{R}_3=3$.
In analogy to the superharmonic criterion for Laughlin two-body 
correlations, their occurrence is related to the form of the 
three-body pseudopotential at short range.
The spectra of a model three-body repulsion are calculated, and 
the zero-energy Moore--Read ground state, its $(\pm e/4)$-charged 
quasiparticles, and the magnetoroton and pair-breaking bands are 
all identified.
The quasiparticles are correctly described by a composite fermion 
model appropriate for Halperin's $p$-type pairing with Laughlin 
correlations between the pairs.
However, the Moore--Read ground state, and specially its excitations, 
have small overlaps with the corresponding Coulomb eigenstates when 
calculated on a sphere.
The reason lies in surface curvature which affects the form of pair 
pseudopotential for which the ``$\mathcal{R}_3\!>\!3$'' three-body 
correlations occur.
In finite systems, such pseudopotential must be slightly superharmonic 
at short range (different from Coulomb pseudopotential). 
However, the connection with the three-body pseudopotential is 
less size-dependent, suggesting that the Moore--Read state and 
its excitations are a more accurate description for experimental 
$\nu={5\over2}$ states than could be expected from previous 
calculations.
\end{abstract}
\maketitle

\section{Introduction}

The fractional quantum Hall (FQH) effect\cite{tsui82,laughlin83} 
is a many-body phenomenon consisting of the quantization of Hall 
conductance and the simultaneous vanishing of longitudinal 
resistance of a high-mobility quasi-two-dimensional electron 
gas at a strong magnetic field $B$ and low density $\varrho$, 
corresponding to certain universal fractional values of the 
Landau level (LL) filling factor $\nu=2\pi\varrho\lambda^2$ 
(where $\lambda=\sqrt{hc/eB}$ is the magnetic length).
This macroscopic phenomenon is a consequence of the formation 
of incompressible liquid ground states (GS's) with quasiparticle 
(QP) excitations.\cite{laughlin83}
It depends on correlations in partially filled degenerate LL's, 
entirely determined by a Haldane pseudopotential\cite{haldane87} 
defined as the pair interaction energy $V_2$ as function of 
relative pair angular momentum $\mathcal{R}_2$.

The Haldane hierarchy\cite{haldane83,laughlin84,halperin84,%
sitko97,hierarchy} of most prominent FQH states in LL$_0$ 
(lowest LL), equivalent to Jain's sequence\cite{jain89} of 
filled composite fermion\cite{lopez91,halperin93} (CF) levels, 
results for Laughlin correlations\cite{halperin83,haldane85,%
rezayi91} (between electrons or QP's) induced by pseudopotentials 
strongly superharmonic\cite{parentage,fivehalf} at short range.
However, the FQH states with different, non-Laughlin correlations 
occur as well. 
E.g., pairing in a half-filled LL$_1$ (first excited LL) is firmly 
established in the $\nu={5\over2}$ state,\cite{willet87,eisenstein88,%
gammel88,pan99} while correlations between CF's in their CF-LL$_1$ 
responsible for the FQH effect\cite{pan03,goldman90} at $\nu={3\over8}$ 
or ${4\over11}$ are not yet completely understood.

The lack of superharmonic behavior of the pseudopotential at short 
range together with the occurrence of clearly non-Laughlin half-filled 
FQH states suggests pairing in both LL$_1$ and CF-LL$_1$.
Proposed trial states include Halperin\cite{halperin83} and 
Haldane--Rezayi\cite{haldane88} states with Laughlin correlations
between spin-triplet and -singlet pairs, respectively, and the 
Moore--Read\cite{moore91,rezayi00} pfaffian state that can be 
defined as a zero-energy ground state of a short-range three-body 
repulsion.\cite{greiter91}
These pair states have all been studied in great detail\cite{wen93,%
milovanovic96,nayak96,read96,read99,gurarie00,tserkovnyak03} because
of their anticipated exotic properties, such as nonabelian QP 
statistics\cite{moore91} or existence of pair-breaking neutral 
fermion excitations.\cite{greiter91}
However, choosing the correct one for specific real FQH systems 
is somewhat problematic.
In the following we concentrate on the half-filled LL$_1$. 
The question of pairing in CF-LL$_1$ is addressed elsewhere.
\cite{qepair1} 

The trouble with the Halperin state\cite{halperin83} is that
because the relative angular momentum of the constituent pairs 
is not a conserved quantity, it is more of an intuitive concept 
for the correlations than a well-defined trial wavefunction 
obeying all required symmetries.
E.g., description of the pair--pair interaction by an effective 
pseudopotential is not rigorous,\cite{qepair1} and the harmonic 
criterion\cite{parentage,fivehalf} that would relate the 
occurrence of Laughlin pair--pair correlations with the 
electron pseudopotential is not exact.
Consequently, it has not been clear what exactly is the model
interaction that induces such correlations (and such ground state).
In fact, it has been (erroneously) assumed\cite{greiter91} 
that this paired state results for pseudopotentials attractive 
at short range rather than harmonically repulsive as in LL$_1$, 
which would suggest that it is not an adequate trial state for 
the $\nu={5\over2}$ FQH effect.

The Moore--Read wavefunction on the other hand is well-defined.
\cite{moore91,rezayi00,greiter91}
However, it only occurs for interactions with very particular
short-range behavior, while the pseudopotentials in realistic 
experimental systems depend on sample parameters like the layer 
width $w$, magnitude and tilt of the magnetic field, etc.
Moreover, finite-size calculations indicate that realistic 
Coulomb pseudopotentials are too weak at short range (by up 
to $\sim10\%$ for $w=0$) to induce a Moore--Read ground state.
\cite{morf98,rezayi00}
This would seem to imply that the Moore--Read state does not 
describe the $\nu={5\over2}$ FQH state quite as accurately as 
a Laughlin state describes the actual $\nu={1\over3}$ ground 
states.
The occurrence of the $\nu={5\over2}$ FQH effect could still be 
attributed to the observation that the calculated excitation 
gaps are much less sensitive to the details of the pseudopotential 
than the wavefunctions.
However, poor accuracy of the Moore--Read wavefunction puts 
doubt on the occurrence of those of its properties in realistic 
$\nu={5\over2}$ systems that depend more critically on the 
correlations.
As these properties (including nonabelian QP's) are so much
more fascinating than plain incompressibility, the question 
of whether they indeed remain only an unrealized theoretical 
concept is quite significant.
Theoretical insight is especially valuable in this problem 
because of the difficulty with direct experimental evidence.
\cite{foster03}

In this paper we report on numerical calculations of 
three-body correlation functions (defined in analogy to 
Haldane pair amplitudes\cite{haldane87}) of the half-filled 
shells with model pair interactions.
We find that the vanishing of the triplet amplitude 
$\mathcal{G}_3(\mathcal{R}_3)$ for the minimum triplet 
relative angular momentum $\mathcal{R}_3=3$, distinctive 
for the Moore--Read state, occurs for the slightly 
superharmonic pseudopotential, different from the nearly 
harmonic one of LL$_1$.
When the vanishing of $\mathcal{G}_3(3)$ is related to
the triplet rather than pair pseudopotential, it becomes
evident that the short-range anharmonicity of the critical 
pair interaction is a finite-size curvature effect on a 
sphere that could disappear in the thermodynamical limit.
Consequently, the three-body correlations defining the 
Moore--Read state and consisting of the avoidance of the 
$\mathcal{R}=3$ hard-core appear to be a much better 
description of the real $\nu={5\over2}$ electron systems 
than expected before.

Based on the anharmonicity of the triplet pseudopotential
we also argue that the Halperin paired state is not an 
adequate model for subharmonic interactions (e.g., in 
CF-LL$_1$) because of the tendency to form larger clusters.
However, the avoidance of $\mathcal{R}_3=3$ triplet states
coinciding with an increased number of $\mathcal{R}_2=1$ 
pairs (compared to the minimum value of a Laughlin-correlated 
state) that occurs for the harmonic interactions is precisely 
the signature of Halperin's pairing concept, which therefore 
is established as a valid model for the Moore--Read state and 
its excitations.
Indeed, the energy spectra of the model three-body repulsion 
show the low-energy bands containing quasielectron (QE) and 
quasihole (QH) excitations of charge $\mathcal{Q}=\pm e/4$, 
in perfect agreement with Halperin's picture for the Laughlin 
state of $\mathcal{R}_2=1$ pairs (with the obvious exception 
being the additional pair-breaking excitation\cite{greiter91}).

\section{Two-body correlations}

\subsection{Haldane pair pseudopotential}

\begin{figure}
\resizebox{3.4in}{1.61in}{\includegraphics{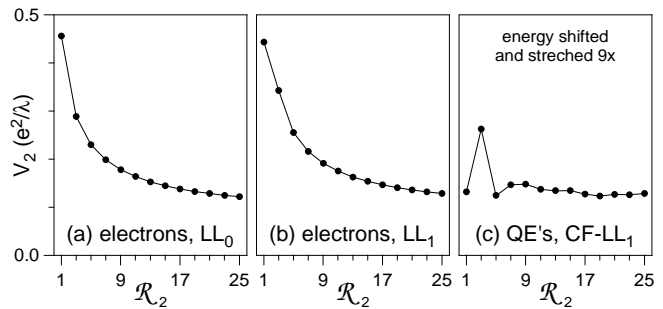}}
\caption{
\label{fig1}
   Pair interaction pseudopotentials (pair interaction energy 
   $V_2$ {\sl vs.}\ relative pair angular momentum 
   $\mathcal{R}_2$) for electrons in the lowest (a) and first 
   excited LL (b), and for QE's of the Laughlin $\nu={1\over3}$ 
   state (c).
   The values of $V_2$ in frame (c) were calculated by Lee {\sl 
   et al.}\cite{lee01} and are only known up to a constant.
   $\lambda$ is the magnetic length.}
\end{figure}

Within a degenerate LL, the many-body Hamiltonian only contains
the interaction term, which is completely determined by the
discrete (Haldane) pseudopotential $V_2(\mathcal{R}_2)$ defined 
as pair interaction energy $V_2$ as a function of relative pair 
angular momentum $\mathcal{R}_2$.
For identical fermions/bosons, $\mathcal{R}_2$ takes on odd/even 
integer values, respectively, and the larger $\mathcal{R}_2$ 
corresponds to a larger average pair separation $\sqrt{\left<
r^2\right>}$.
On a sphere, $\mathcal{R}_2=2l-L_2$ where $l$ is the single-particle 
angular momentum of the shell (LL), and $L_2$ is the total pair 
angular momentum.
(We use the following standard notation for Haldane\cite{haldane83} 
spherical geometry: $l=Q+n$ for the $n$th LL, $2Q=4\pi R^2B/\phi_0$ 
is the magnetic monopole strength $\phi_0=hc/e$ is the flux quantum,
$R$ is the sphere radius, and $\lambda=R/\sqrt{Q}$ is the magnetic
length.)
Importantly, $V_2(\mathcal{R}_2)$ combines information about both 
interaction potential $V(r)$ and the single-particle wavefunctions 
allowed within the Hilbert space restricted to a LL.
The pseudopotentials obtained for the electrons in LL$_0$ and LL$_1$, 
and for Laughlin QE's in CF-LL$_1$ are shown in Fig.~\ref{fig1}.

\subsection{Pair amplitudes}

\begin{figure}
\resizebox{3.4in}{3.40in}{\includegraphics{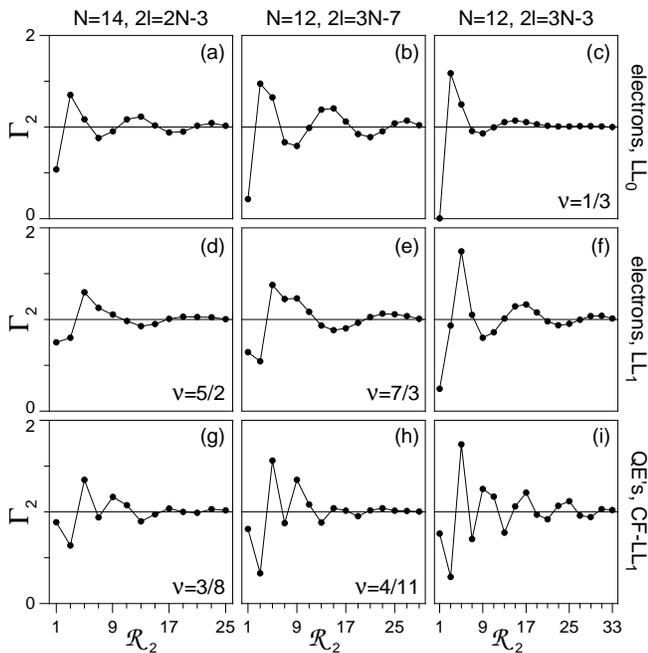}}
\caption{
\label{fig2}
   Pair-correlation functions (pair amplitude $\Gamma_2$ 
   {\sl vs.}\ relative pair angular momentum $\mathcal{R}_2$)
   calculated on a sphere for the lowest $L=0$ states of $N$ 
   particles interacting through pseudopotentials shown in 
   Fig.~\ref{fig1}, at values of $2l$ corresponding to different 
   FQH states at filling factors $\nu={1\over2}$ and ${1\over3}$.}
\end{figure}

The pair correlations induced by a specific $V_2(\mathcal{R}_2)$ 
are conveniently described by a discrete pair amplitude function 
$\mathcal{G}_2(\mathcal{R}_2)$, defined as the number of pairs 
$\mathcal{N}_2$ with a given $\mathcal{R}_2$ divided by the 
total pair number,
\begin{equation}
   \mathcal{G}_2(\mathcal{R}_2)
   ={N\choose2}^{-1}\mathcal{N}_2(\mathcal{R}_2).
\end{equation}
It immediately follows from the expression for the total interaction 
energy of an $N$-body state,
\begin{equation}
\label{eqE2}
   E={N\choose2}\sum_{\mathcal{R}_2}
   \mathcal{G}_2(\mathcal{R}_2)\,V_2(\mathcal{R}_2),
\end{equation} 
that the low-energy many-body states generally have a large/small 
amplitude at those values of $\mathcal{R}_2$ corresponding to 
small/large repulsion $V_2(\mathcal{R}_2)$.
In Fig.~\ref{fig2} we compare the pair amplitudes obtained in Haldane 
spherical geometry for $N=12$ and 14 particles confined in angular 
momentum shells with degeneracy $g=2l+1$ corresponding to the filling 
factors $\nu\sim{1\over3}$ and ${1\over2}$ and interacting through 
the pseudopotentials of Fig.~\ref{fig1}.
Although for each system $(V_2,N,g)$ we only show the data for the 
lowest $L=0$ state, virtually identical $\mathcal{G}_2(\mathcal{R}_2)$
functions are obtained for all low-energy states of each system.

The chosen values of $2l$ and $N$ correspond to three different 
sequences of finite-size spherical systems known to represent 
the following FQH states observed experimentally on a plane.
The $2l=2N-3$ sequence describes the paired Moore--Read\cite{moore91} 
$\nu={1\over2}$ state in LL$_1$ (corresponding to the total electron 
filling factor $\nu={5\over2}$) and the $\nu={1\over2}$ state of 
QE's in CF-LL$_1$ identified numerically\cite{qepair1} for $N=6$, 
10, and 14, and corresponding to the FQH effect\cite{pan03} at 
$\nu={3\over8}$.
The $2l=3N-7$ sequence describes the (not well understood) $\nu=
{1\over3}$ states in both LL$_1$\cite{fivehalf} and CF-LL$_1$,
\cite{qepair1} corresponding to the $\nu={7\over3}$ state
\cite{willet87,eisenstein88,gammel88,pan99} and $\nu={4\over11}$ 
state,\cite{pan03,goldman90} respectively.
Finally, the $2l=3N-3$ sequence describes the Laughlin
\cite{laughlin83} $\nu={1\over3}$ state in LL$_0$.

The pair amplitude calculated for a completely filled shell 
(the $\nu=1$ state) with a given $2l$ is a decreasing straight
line,
\begin{equation}
   \mathcal{G}_2^{\rm full}(\mathcal{R}_2)
   ={4l+1-2\mathcal{R}_2\over l(2l+1)},
\end{equation}
which is a finite-size edge effect.
In the $2l\rightarrow\infty$ limit corresponding to an infinite
plane, $\mathcal{N}_2^{\rm full}(\mathcal{R}_2)=N$ and the 
ratio $\mathcal{N}_2^{\rm full}(\mathcal{R}_2)/N\equiv
\Gamma^{\rm full}(\mathcal{R}_2)=1$ is the appropriately 
renormalized pair amplitude in this geometry.

The overall linear decrease of $\mathcal{G}_2(\mathcal{R}_2)$
appears also at $\nu<1$, and it should be ignored
in the analysis of correlations.
Therefore, in Fig.~\ref{fig2} we actually plot 
\begin{equation}
   \Gamma_2(\mathcal{R}_2)
   =1+{\mathcal{G}_2(\mathcal{R}_2)
      -\mathcal{G}_2^{\rm full}(\mathcal{R}_2)\over
       \mathcal{G}_2^{\rm full}(1)},
\end{equation}
in which the linear decrease is eliminated and the scaling 
appropriate for an infinite plane is used, to ensure that
$\Gamma_2(1)\propto\mathcal{G}_2(1)$, that $\Gamma_2
(\mathcal{R}_2)=1$ for finite-size $\nu=1$ states, and that 
$\Gamma_2(\mathcal{R}_2)$ converges to the pair-correlation 
function on the plane when $N$ is increased.

In all frames, $\Gamma_2$ is significantly different from 1 
only at small $\mathcal{R}_2$, and the oscillations around 
this value quickly decay beyond $\mathcal{R}_2\sim7$.
This can be interpreted as a short correlation range $\xi$ 
in all studied systems and it justifies the use of finite-size 
calculations (requiring that $\xi\ll R$)

Clearly, three different interactions result in quite different 
correlations.
In LL$_0$ (a--c), the dominant tendency is the avoidance of 
$\mathcal{R}_2=1$ (Laughlin correlations) at a cost of having 
a large number of pairs with $\mathcal{R}_2=3$.
Around half-filling of LL$_1$ (d,e), the numbers of pairs 
with $\mathcal{R}_2=1$ and 3 are about equal and both small.
Finally, in a partially filled CF-LL$_1$ (g--i), the 
$\mathcal{R}_2=3$ pair state is maximally avoided.

\subsection{Model interaction and pair-correlation regimes}

The fact that $\Gamma_2\approx1$ at long range explains also 
why the low-energy wavefunctions are virtually insensitive to 
the exact form of $V_2(\mathcal{R}_2)$ beyond a few leading 
parameters at $\mathcal{R}_2=1$, 3, \dots.
Furthermore, due to the sum rules obeyed by pair amplitudes, 
\cite{parentage,fivehalf,sum-rules} the harmonic pseudopotentials 
$V_2^{\rm H}(\mathcal{R}_2)=c_0-c_1\mathcal{R}$ (with constant 
$c_0$ and $c_1$) induce no correlations, and only the anharmonic 
contributions to $V_2(\mathcal{R}_2)$ at small $\mathcal{R}_2$ 
(short range) affect the pair-correlation functions.
Indeed, simple model pseudopotentials with only two nonvanishing 
leading parameters are known\cite{greiter91,fivehalf,qepair1} 
to accurately reproduce correlations 
shown in Fig.~\ref{fig2}.
Let us define such $U_\alpha(\mathcal{R}_2)$ with 
\begin{eqnarray}
\label{eqU}
   U_\alpha(1)&=&1-\alpha,\nonumber\\
   U_\alpha(3)&=&\alpha/2.
\end{eqnarray}
$U_0$ and $U_1$ are the two extremal pseudopotentials with only
one anharmonic term.
$U_{1\over2}$ is harmonic through $\mathcal{R}_2=1$, 3, and 5, and 
thus it favors equally the avoidance of both $\mathcal{R}_2=1$ and 
3 pairs (any distribution of the pair amplitude between the
$\mathcal{R}_2=1$, 3, and 5 states that satisfies the sum rules
yields the same total energy $E$).

\begin{figure}
\resizebox{3.4in}{1.58in}{\includegraphics{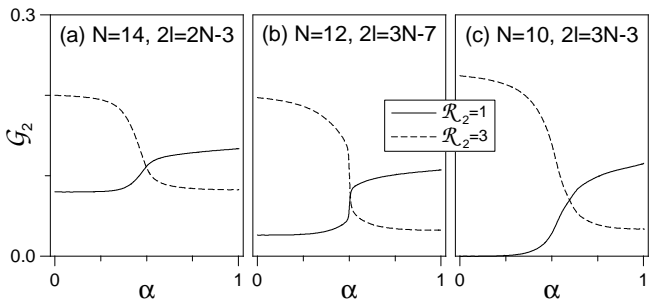}}
\caption{
\label{fig3}
   Dependence of pair amplitudes $\mathcal{G}_2$ on parameter 
   $\alpha$ of pair interaction $U_\alpha$ defined by 
   Eq.~(\ref{eqU}), calculated on a sphere for the lowest $L=0$ 
   states of $N$-particle systems representing the same FQH 
   states as used in Fig.~\ref{fig2}.}
\end{figure}

In Fig.~\ref{fig3} we plot $\mathcal{G}_2(1)$ and $\mathcal{G}_2(3)$ 
as a function of $\alpha$ for the lowest $L=0$ state in three 
finite-size systems representing the same series of FQH states
as used in Fig.~\ref{fig2}.
The correlations in a partially filled LL$_0$ (Laughlin correlations), 
LL$_1$, and CF-LL$_1$ are well reproduced by $U_\alpha$ with $\alpha
\approx0$, ${1\over2}$, and 1, respectively.
The correlations at $\alpha=0$ (i.e., in LL$_0$) and at $\alpha=1$ 
(i.e., in CF-LL$_1$) can easily be expressed in terms of pair 
amplitudes.
With $U(1)$ or $U(3)$ being the only nonvanishing (and positive) 
coefficient, it follows from Eq.~(\ref{eqE2}) that the low-energy 
states must have the minimum allowed (within the available Hilbert
space) $\mathcal{G}_2(1)$ or $\mathcal{G}_2(3)$, respectively.

In case of Laughlin correlations, because of the simple form of 
single-particle wavefunctions in LL$_0$, the complete avoidance 
of the $\mathcal{R}_2=1$ pairs (possible at $\nu\le{1\over3}$) 
appears in form of a Jastrow factor in the Laughlin wavefunction.
It justifies the mean-field CF picture that essentially attributes 
the reduction of the many-body degeneracy caused by the 
$\mathcal{R}_2=1$ hard-core (or ``correlation hole'') to an 
effective, reduced magnetic field.

For QE's, the tendency to have small $\mathcal{G}_2(3)$ and,
consequently, significant $\mathcal{G}_2(1)$ (compared to a 
Laughlin-correlated state at the same $\nu$) has been
interpreted as $\mathcal{R}_2=1$ pairing (although most recent 
numerical studies are not conclusive about how the pairs 
correlate with one another, and thus the question of the 
origin of the excitation gap observed at $\nu={3\over8}$ 
or ${4\over11}$ remains open).

In a partially filled LL$_1$ the situation is more complicated.
Because $V_2(\mathcal{R}_2)$ is nearly harmonic at short range 
($\alpha\sim{1\over2}$), the energy is nearly independent of 
the relative occupation of the $\mathcal{R}_2=1$ and 3 pair 
states.
Therefore, the correlations cannot be easily expressed in terms 
of pair amplitudes (although the linear combination of 
$\mathcal{G}_2(1)$ and $\mathcal{G}_2(3)$ equal to the total 
energy $E$ is obviously minimized at its corresponding value 
of $\alpha$).
However, it turns out that it is the short-range three-body 
correlations that determine the low-energy states in this regime.
Soon after its introduction, the half-filled Moore--Read state 
was shown\cite{greiter91} to be an exact zero-energy eigenstate 
of a model short-range three-body repulsion, and the spectra of 
this interaction were later studied in detail.\cite{wen93,read96}
Below we analyze the three-body correlations directly, by
the calculation of an appropriate correlation function.

\section{Three-body correlations}

\subsection{Three-body pseudopotential}

In analogy to the avoidance of the strongly repulsive pair 
states, the three-body states with sufficiently high energy 
(compared to the rest of the three-body spectrum) will also 
be avoided in the low-energy many-body states.
For the pairs, the eigenstates are uniquely labeled by 
$\mathcal{R}_2$, and the criterion for the avoidance of 
a specific $\mathcal{R}_2$ is\cite{parentage,fivehalf}
that it corresponds to the dominant positive anharmonic 
term of $V_2(\mathcal{R}_2)$.
The three-body states are also labeled by the relative (with 
respect to the center of mass) angular momentum $\mathcal{R}_3$.
The allowed values are $\mathcal{R}_3=3$ or $\mathcal{R}_3\ge5$,
and larger $\mathcal{R}_3$ means larger expectation value of 
the area spanned by the three particles.\cite{hawrylak95}
On a sphere, $\mathcal{R}_3=3l-L_3$, where $L_3$ is the total 
triplet angular momentum.

\begin{figure}
\resizebox{3.4in}{3.26in}{\includegraphics{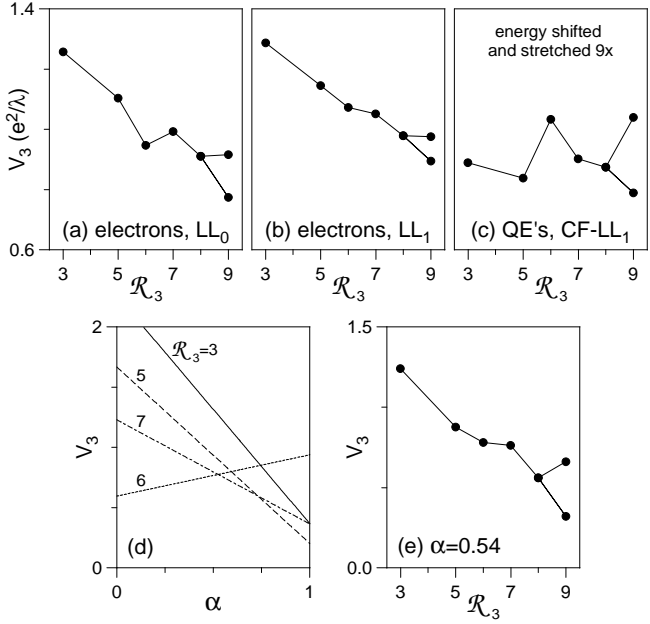}}
\caption{
\label{fig4}
   (a,b,c)
   Triplet interaction pseudopotentials (triplet interaction 
   energy $V_3$ {\sl vs.}\ relative triplet angular momentum 
   $\mathcal{R}_3$) for pair pseudopotentials shown in 
   Fig.~\ref{fig1}.
   $\lambda$ is the magnetic length.
   (d) 
   Dependence of coefficients $V_3$ on parameter $\alpha$ of 
   pair interaction  $U_\alpha$ defined by Eq.~(\ref{eqU}).
   (e) 
   Pseudopotential $V_3(\mathcal{R}_3)$ calculated for pair 
   interaction $U_{0.54}$.}
\end{figure}

Since no degeneracies appear in the $V_3(\mathcal{R}_3)$ energy 
spectrum for $\mathcal{R}_3<9$, its low-$\mathcal{R}_3$ part can 
be considered a three-body pseudopotential analogous to 
$V_2(\mathcal{R}_2)$.
The three-body pseudopotentials $V_3(\mathcal{R}_3)$ obtained for 
different pair pseudopotentials $V_2(\mathcal{R}_2)$ of Fig.~\ref{fig1} 
are shown in the upper frames of Fig.~\ref{fig4}.
The nonmonotonic behavior of $V_3(\mathcal{R}_3)$ in frame (c) 
most likely precludes the tendency to avoid the $\mathcal{R}_3=3$ 
triplet state in QE systems.
On the other hand, it seems plausible that the monotonic character 
of $V_3(\mathcal{R}_3)$ in frame (b) might lead to the avoidance
of the same $\mathcal{R}_3=3$ triplet state in a partially filled 
LL$_1$.

The dependence of $V_3(\mathcal{R}_3)$ on $V_2(\mathcal{R}_2)$ can
be captured by plotting the leading $V_3$ coefficients as a function
of parameter $\alpha$ of the model pair pseudopotential $U_\alpha$,
as shown in frame (d).
For $\mathcal{R}_3<9$ the triplet wavefunctions are fixed and so 
are their $\mathcal{G}_2$ amplitudes, and hence the dependences 
$V_3(\alpha)$ are all linear.
Only around $\alpha\sim{1\over2}$ is the $V_3(\mathcal{R}_3)$ 
function superlinear for small $\mathcal{R}_3$, as shown on an 
example for $\alpha=0.54$ in frame (e).

\subsection{Three-body amplitudes}

In order to test the hypothesis of the avoidance of the 
$\mathcal{R}_3=3$ triplet eigenstate in partially filled LL$_1$, 
we introduce ``triplet amplitude'' $\mathcal{G}_3(\mathcal{R}_3)$.
It is defined in analogy to the pair amplitude, as an expectation 
value of the operator $\hat\mathcal{P}_{ijk}(\mathcal{R}_3,\beta_3)$ 
projecting a many-body state $\Psi$ onto the subspace in which the 
three particles $ijk$ are in an eigenstate $\left|\mathcal{R}_3,
\beta_3\right>$ (here, $\beta_3$ is an additional index to distinguish 
degenerate multiplets at the same $\mathcal{R}_3$; it can be 
omitted for $\mathcal{R}_3<9$).
The interaction Hamiltonian written in a three-body form using 
$\hat\mathcal{P}_{ijk}$ reads
\begin{equation}
   \hat\mathcal{H}=\sum_{i<j<k}\sum_{\mathcal{R}_3,\beta_3}
   \hat\mathcal{P}_{ijk}(\mathcal{R}_3,\beta_3)\,V(\mathcal{R}_3,\beta_3).
\end{equation} 
The triplet amplitude is
\begin{equation}
   \mathcal{G}(\mathcal{R}_3,\beta_3)=
   {N\choose3}^{-1}
   \left<\Psi\right|
   \sum_{i<j<k}\hat\mathcal{P}_{ijk}(\mathcal{R}_3,\beta_3)
   \left|\Psi\right>,
\end{equation}
which for a totally antisymmetric $\Psi$ is equivalent to
\begin{equation}
   \mathcal{G}(\mathcal{R}_3,\beta_3)=
   \left<\Psi\right|
   \hat\mathcal{P}_{123}(\mathcal{R}_3,\beta_3)
   \left|\Psi\right>.
\end{equation}
Pair amplitudes defined in this way are normalized to
\begin{equation}
   \sum_{\mathcal{R}_3,\beta_3}
   \mathcal{G}_3(\mathcal{R}_3,\beta_3)=1,
\end{equation}
so that they measure the fraction of all triplets being in 
a given eigenstate,
\begin{equation}
   \mathcal{G}_3(\mathcal{R}_3,\beta_3)
   ={N\choose3}^{-1}\mathcal{N}_3(\mathcal{R}_3,\beta_3).
\end{equation}
The energy of $\Psi$ is expressed as
\begin{equation}
\label{eqE3}
   E={N\choose3}\sum_{\mathcal{R}_3,\beta_3}
   \mathcal{G}_3(\mathcal{R}_3,\beta_3)\,V_3(\mathcal{R}_3,\beta_3).
\end{equation}
On a sphere, triplet amplitudes are connected with the third-order 
parentage coefficients\cite{cowan81} $G_3(L_3,\beta_3;L_3',\beta_3')$, 
i.e., the expansion coefficients of a totally antisymmetric state 
$\Psi$ in a basis of product states in which particles $(1,2,3)$ and 
$(4,5,\dots,N)$ are in the 3- and $(N-3)$-body eigenstates 
$\left|L_3,\beta_3\right>$ and $\left|L_3',\beta_3'\right>$, 
respectively,
\begin{equation}
   \mathcal{G}_3(L_3,\beta_3)=
   \sum_{L_3',\beta_3'}|G_3(L_3,\beta_3;L_3',\beta_3')|^2.
\end{equation}
Note that to obey standard notation for parentage coefficients,
in the above equation we use total angular momentum $L_3$ instead 
of the relative one $\mathcal{R}_3=3l-L_3$ to label triplet states.
Also, we omit index $\Psi$ in $\mathcal{G}_3$ and $G_3$.

The following operator identity\cite{sum-rules} 
\begin{equation}
   \hat{L}^2+N(N-2)\,\hat{l}^2=\sum_{i<j}\hat{L}_{ij}^2
\end{equation}
connects the total $N$-body angular momentum ($L$) with the 
single-particle and pair angular momenta $l$ and $L_{ij}$.
We used it earlier to show that harmonic pair pseudopotentials 
cause no correlations.
It can be generalized to the following form
\begin{equation}
   \hat{L}^2+{N(N-K)\over K-1}\,\hat{l}^2
   ={N-2\choose K-2}^{-1}\!\!\!\!\!
   \sum_{i_1<\dots<i_K}\!\!\!\hat{L}_{i_1\dots i_K}^2,
\end{equation}
By taking the expectation values of both sides of the above
equation in the (totally antisymmetric) state $\Psi$ and
using the expansion of $\Psi$ in terms of the $K$th-order 
parentage coefficients we obtain
\begin{eqnarray}
\label{eqSUM}
   &&L(L+1)+{N(N-K)\over K-1}\,l(l+1)\\\nonumber
   &&={N(N-1)\over K(K-1)} 
   \sum_{L_K,\beta_K} \mathcal{G}_K(L_K,\beta_K)\,L_K(L_K+1),
\end{eqnarray}
an additional (besides normalization) sum rule obeyed
by the amplitudes $\mathcal{G}_K$.

Just as for the specific $K=2$ case discussed earlier,
\cite{sum-rules} the above sum rule (\ref{eqSUM}) together 
with an appropriate version of Eq.~(\ref{eqE2}) or (\ref{eqE3}) 
immediately implies that if the $K$-body interaction 
pseudopotential $V_K$ is linear in $L_K(L_K+1)$, all 
$N$-body multiplets with the same $L$ are degenerate.
In the limit of infinite LL degeneracy $g=2l+1$ corresponding
to an infinite sphere radius (vanishing curvature) i.e., to
the planar geometry, the linearity in $L_K(L_K+1)$ translates
into the linearity in $\mathcal{R}_K=Kl-L_K$, and it turns
out that the linear part of $V_K(\mathcal{R}_K)$ causes no
correlations.

\subsection{Three-body correlation hole}

\begin{figure}
\resizebox{3.4in}{1.54in}{\includegraphics{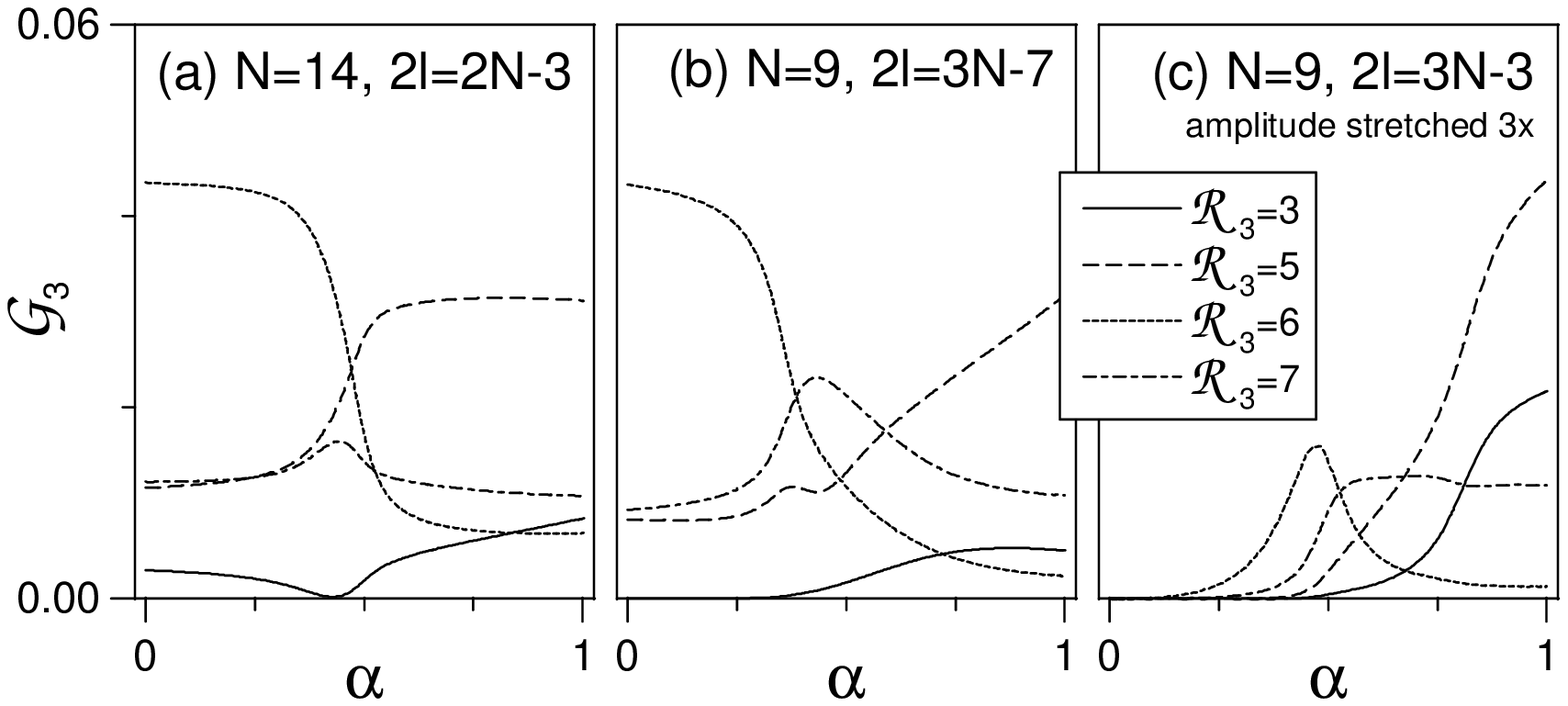}}
\caption{
\label{fig5}
   Dependence of triplet amplitudes $\mathcal{G}_3$ on 
   parameter $\alpha$ of pair interaction $U_\alpha$ defined 
   by Eq.~(\ref{eqU}), calculated on a sphere for the lowest 
   $L=0$ states of $N$-particle systems representing the same 
   FQH states as used in Figs.~\ref{fig2} and \ref{fig3}.}
\end{figure}

Let us now turn back to the numerical results. 
In Fig.~\ref{fig5} we plot the dependence of the leading 
$\mathcal{G}_3(\mathcal{R}_3)$ coefficients on $\alpha$, 
calculated in the lowest $L=0$ state of three different 
systems belonging to the same sequences of finite-size FQH 
states as used earlier in Figs.~\ref{fig2} and \ref{fig3}.
Clearly, all triplet amplitudes significantly depend on 
$\alpha$, but we especially want to point out the following 
three features for $\mathcal{R}_3=3$:
(i) the tendency to avoid $\mathcal{R}_2=1$ pairs at 
$\alpha\sim0$ is {\em not} synonymous with the avoidance 
of $\mathcal{R}_3=3$ triplets at $\nu={1\over2}$,
(ii) $\mathcal{G}_3(3)$ {\em vanishes} for $\alpha\approx
{1\over2}$ at $\nu={1\over2}$.
(iii) $\mathcal{G}_3(3)$ {\em increases} when $\alpha$ 
increases beyond ${1\over2}$ in all frames.

Before we concentrate on the Moore--Read state, let us 
note that observation (iii) confirms the suspicion based 
on the form of triplet pseudopotential $V_3(\mathcal{R}_3)$ 
of Fig.~\ref{fig4}(c) that (against an earlier assumption
\cite{greiter91}) the Halperin paired state\cite{halperin83} 
is not an adequate description for systems with subharmonic 
pseudopotentials at short range.
In particular (against our earlier expectation\cite{qepair2} 
but in agreement with our later numerical results\cite{qepair1}) 
such model appears inappropriate for the QE's in CF-LL$_1$ 
at $\nu={1\over2}$ or ${1\over3}$, corresponding to the FQH 
states at $\nu={3\over8}$ and ${4\over11}$.
Instead of Halperin's paring, grouping of pairs into larger 
clusters seems to occur for the QE's, although we are not able 
to define their correlations more specifically.

\begin{figure}
\resizebox{3.4in}{2.19in}{\includegraphics{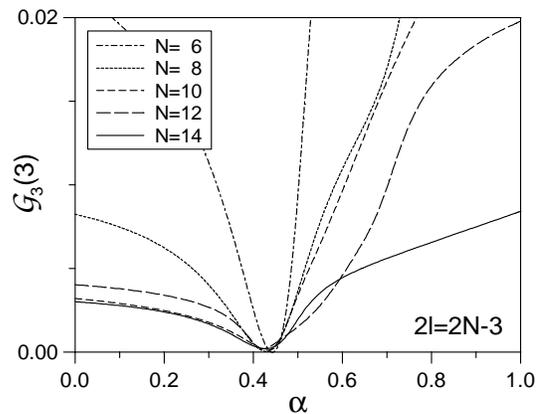}}
\caption{
\label{fig6}
   The $\mathcal{G}_3(3)$ {\sl vs.}\ $\alpha$ curve for 
   $\nu={1\over2}$ shown in Fig.~\ref{fig5}(a), magnified 
   and replotted for different particle numbers $N$.}
\end{figure}

Let us now discuss observations (i) and (iii) in more detail.
In Fig.~\ref{fig6} we plot $\mathcal{G}_3(3)$ as a function 
of $\alpha$ for $N=6$ to 14 (only even values, because the 
Moore--Read state at $2l=2N-3$ is a paired state).
For each $N$, $\mathcal{G}_3(3)$ drops to essentially a zero 
at exactly $\alpha_0\approx{1\over2}$, but it increases 
quickly when $\alpha$ moves away from this critical value.
This result is consistent with the calculations of overlaps 
of the exact ground states of modified Coulomb interaction 
with the exact Moore--Read trial state.\cite{rezayi00,morf98}

It is difficult to reliably extrapolate the values of $\alpha_0$ 
obtained from Fig.~\ref{fig6} to an infinite (planar) system.
However, we notice the following connection with Fig.~\ref{fig4}(d)
that depends on $2l$ much more regularly.
The pair amplitudes $\mathcal{G}_2=[\mathcal{G}_2(1),\mathcal{G}_2
(3),\dots]$ of the $\mathcal{R}_3<9$ triplets can be calculated. 
On a sphere, they slightly depend on $2l$ (on curvature), but the 
values appropriate for a plane (the $g\rightarrow\infty$ limit) 
are: $[{3\over4},{1\over4}]$, $[{9\over16},{1\over8},{5\over16}]$, 
and $[{3\over16},{5\over8},{3\over16}]$ for $\mathcal{R}_3=3$, 5, 
and 6, respectively.
Using these values and Eq.~(\ref{eqE2}) one can determine the
range $\alpha$ over which $V_3(\mathcal{R}_3)$ is superlinear
at short range.
The requirement that ${1\over2}[V_3(3)-V_3(5)]>V_3(5)-V_3(6)$
and $V_3(5)-V_3(6)>V_3(6)-V_3(7)$ limits $\alpha$ to a rather
narrow window of approximately
\begin{equation}
\label{eqLIM}
   0.5<\alpha+{1\over4l}<0.58.
\end{equation}
For a reason we do not completely understand (but that is 
connected with a neglected and complicated behavior of 
$V_3(\mathcal{R}_3)$ at $\mathcal{R}_3>7$), the value of 
$\alpha_0$ in finite systems (see Fig.~\ref{fig6}) is much 
closer to the lower limit of Eq.~(\ref{eqLIM}).
Therefore, we expect that $\alpha_0$ will follow this lower 
limit with increasing $2l$, and the value appropriate 
for a planar system should be even closer to  ${1\over2}$ 
than the finite-size results of Fig.~\ref{fig6}.
And since $U_{1\over2}$ accurately models Coulomb interaction 
in LL$_1$, we conclude that the ``$\mathcal{R}_3>3$'' correlations 
must be an accurate description for experimental $\nu={5\over2}$ 
FQH state (even in narrow samples).
This conclusion is quite different from an earlier discussion
of finite-size numerical wavefunctions\cite{rezayi00,morf98} 
which seemed to imply that a $\sim10\%$ short-range enhancement 
of the Coulomb pseudopotential calculated for $w=0$ in LL$_1$ 
is needed to reach good overlap with the Moore--Read state.

\section{Energy spectra of short-range three-body repulsion}

Knowing that what defines the Moore--Read state is that electrons 
in ${1\over2}$-filled LL$_1$ completely avoid the $\mathcal{R}_3=3$ 
triplet state,\cite{greiter91} let us discuss the energy spectra 
of the model short-range three-body repulsion
\begin{equation}
\label{eqW}
   W(\mathcal{R}_3)=\delta_{\mathcal{R}_3,3}
\end{equation}
which induces precisely this type of correlations.
Similar calculations for slightly smaller systems were earlier 
carried out by Wen\cite{wen93} and by Read and Rezayi.
\cite{read96} 

The three-body interaction matrix elements needed for 
diagonalization in the configuration interaction (CI) 
basis are connected with the triplet spectrum 
$V_3(\mathcal{R}_3,\beta_3)$ through expansion parameters 
$C^A_B\equiv\left<A|B\right>$ analogous to the pair 
Clebsch-Gordan coefficients,
\begin{eqnarray}
   &&\left<m_1,m_2,m_3\right|V_3\left|m_4,m_5,m_6\right>
   \\\nonumber
   &&=\sum_{\mathcal{R}_3,\beta_3}
   C_{m_1,m_2,m_3}^{\mathcal{R}_3,\beta_3*}
   C_{m_4,m_5,m_6}^{\mathcal{R}_3,\beta_3}
   V_3(\mathcal{R}_3,\beta_3).
\end{eqnarray}
For $V_3=W$ the above formula reduces to just one term.
However, diagonalization of $V_3$ is still far more difficult 
than of a (also $L$-conserving) $V_2$ because of a larger number 
of nonzero CI matrix elements (by over 10 times in the systems 
discussed further in this section).

\subsection{Laughlin-like $\nu={1\over2}$ incompressible 
            ground state and excited magnetoroton band}

\begin{figure}
\resizebox{3.4in}{1.50in}{\includegraphics{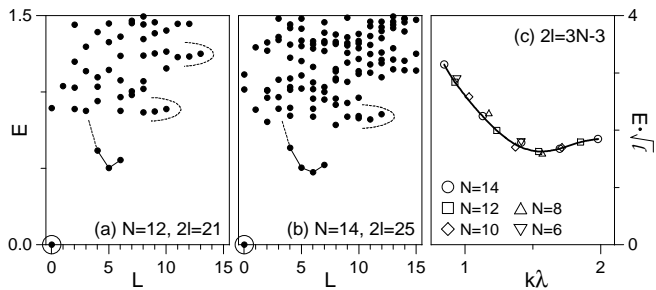}}
\caption{
\label{fig7}
   (a,b)
   Energy spectra (total interaction energy $E$ {\sl vs.}\
   total angular momentum $L$) calculated on a sphere for 
   even numbers of particles $N$ interacting through triplet 
   pseudopotential defined by Eq.~(\ref{eqW}), at the values 
   of $2l=2N-3$ corresponding to the $L=0$ Moore--Read ground 
   state.
   (c) 
   Energy dispersion (excitation energy $E$ as a function of
   wavevector $k$) for the excited magnetoroton band.
   $\lambda$ is the magnetic length.
   Similar results were first obtained by Read and Rezayi
   \cite{read96}.}
\end{figure}

In Fig.~\ref{fig7}(a) and (b) we present the results for $N=12$ 
and 14 and $2l=2N-3$.
As seen on these two examples, for even values of $N$ and for 
$2l=2N-3$ there is exactly one state in the spectrum with 
$E=0$, i.e., with no triplet amplitude at $\mathcal{R}_3=3$.
In other words, the Hilbert subspace with $\mathcal{R}_3>3$
for all triplets contains exactly one state in this case.
At $2l<2N-3$, all states have amplitude at $\mathcal{R}_3=3$,
and at $2l>2N-3$ there is more than one such state.
For odd values of $N$, no such states occur for at $2l\le2N-3$, 
and at $2l>2N-3$ there are always more than one.
This fact makes the Moore--Read yet another beautiful extension 
of the Laughlin idea for the $\nu={1\over3}$ state at $2l=3N-3$ 
being the only state in its Hilbert space with no pair amplitude 
at $\mathcal{R}_2=1$.
Just as the avoidance of more than one pair state generated the 
whole $\nu={1\over3}$, ${1\over5}$, \dots\ sequence, the avoidance
of not just pairs, but triplets (or $K$-body states) gives rise 
to incompressibility at new values of $\nu$.

The analogy to the Laughlin $\nu={1\over3}$ state goes beyond 
the incompressible ground state.
The low-energy excitations clearly form a band that resembles 
the magnetoroton curve.\cite{read96}
In frame (c) we overlay data obtained for different $N=6$ to 14
and plotted as a function of wavevector $k$ (the charge-neutral
excitations carrying $L>0$ on a sphere move along great circles 
of radius $R$, but on a plane they would move along straight lines 
with $k=L/R$).
The continuous character of this band and the minimum at $k\approx
1.5\,\lambda^{-1}$ (very close to $k\approx1.4\,\lambda^{-1}$ of the 
Laughlin $\nu={1\over3}$ state) are clearly visible.

\subsection{Pairing and Laughlin pair--pair correlations}

Before we move on to the spectra at $2l\ne2N-3$ in search of the 
elementary charge excitations of the Moore--Read state, let us 
recall Halperin's\cite{halperin83} concept of Laughlin states 
of $\mathcal{R}_2=1$ pairs that we have also used earlier for 
the half-filling of both LL$_1$\cite{fivehalf} and CF-LL$_1$.
\cite{qepair1,qepair2} 
The increase of $\mathcal{G}_2(1)$ compared to a Laughlin-correlated 
state at the same $\nu$ visible in Fig.~\ref{fig3}(a) can be thought
of as pairing for both $\alpha\sim{1\over2}$ and 1.
However, whether the $\mathcal{R}_1$ pairs will keep far apart from 
one another by avoiding small values of their relative (pair--pair) 
angular momentum (what we would consider Laughlin correlations among
the pairs) has not been established in neither LL$_1$ nor CF-LL$_1$.
Actually, the fact that only for $N=6$, 10, 14, \dots\ (and not for
$N=8$ or 12) do the $L=0$ ground states occur in CF-LL$_1$ suggests
that Halperin's idea could not be correct for the interacting QE's.
However, for the half-filled LL$_1$, the occurrence of a large value
of $\mathcal{G}_2(1)$ and, {\em at the same time}, the vanishing of 
$\mathcal{G}_3(3)$ finally offers support for this idea in the
Moore--Read state.
By effectively acting like a short-range three-body repulsion $W$, 
Coulomb repulsion in LL$_1$ allows grouping electrons into pairs 
(at $\nu$ as large as ${1\over2}$), but it prevents the third 
electron from getting too close to a pair.
As a result, the pairs exist but each pair attains a hard-core that 
results in Laughlin correlation with all other pairs (or unpaired 
electrons), and that can be modeled by a fictitious flux attachment 
in a standard way.

Let us demonstrate how does this picture works for the spectra in 
Fig.~\ref{fig7}.
As a result of the appropriate CF transformation,\cite{fivehalf,%
qepair1,qepair2} $N$ electrons at $2l=(2N+3)\pm\Delta$ are 
converted to $N_2={1\over2}N$ CF's about exactly filling their 
effective CF-LL$_0$ shell with $2l_0^*=2(2l-1)-7(N_2-1)=(N_2-1)
\pm2\Delta$, i.e., with the effective degeneracy 
\begin{equation}
\label{eqEffg}
   g_0^*=N_2\pm2\Delta.
\end{equation}
These CF's correspond to the $\mathcal{R}_2=1$ pairs of electrons,
and their effective angular momentum $l_0^*$ is obtained from $L_2
=2l-1$ by attachment of 7 flux quanta to each pair (4 to account 
for the pair-pair hard-core due to Pauli exclusion principle, 4 to 
model pair--pair Laughlin correlations, and 1 in the opposite 
direction to convert the pairs to fermions).
At exactly $2l=2N-3$, the $N$-body (Moore--Read) ground state is 
equivalent to a full CF-LL$_0$ with $l_0^*={1\over2}(N_2-1)$, i.e., 
to a Laughlin state of $N_2$ pairs.
The magnetoroton band describes QE--QH pair states, with one CF 
excited from the full CF-LL$_0$ to the empty CF-LL$_1$ with 
$l_1^*=l_0^*+1={1\over2}(N_2+1)$.
This band extends up to $L=l_0^*+l_1^*=N_2$.
Higher states above the magnetoroton band contain additional
QE--QH pairs, and the characteristic steps are clearly visible 
in the energy spectra in Fig.~\ref{fig7} (e.g., at $L=(2l_0^*-1)
+(2l_1^*-1)=N-2$ for two QE--QH pairs).

\subsection{Quasiparticles}

\begin{figure}
\resizebox{3.4in}{1.50in}{\includegraphics{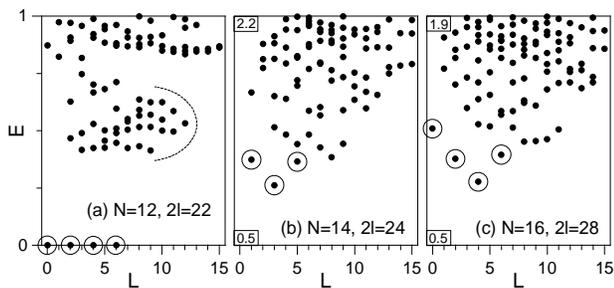}}
\caption{
\label{fig8}
   The same as Fig.~\ref{fig7} but for $2l=2N-2$ (a) and $2l=2N-4$ 
   (b,c) corresponding to two QH's and two QE's in the Moore--Read 
   state, respectively.}
\end{figure}

In Fig.~\ref{fig8} we present sample spectra obtained for even 
values of $N$ and $2l=(2N-3)\pm1$.
At $2l=(2N-3)+1$, there is always a band\cite{wen93,read96} of 
$E=0$ states at $L=N_2$, $N_2-2$, \dots, corresponding to two QH's 
in CF-LL$_0$ of degeneracy $g_0^*=N_2+2$.
This is shown in frame (a) for $N=12$.
Unlike for Laughlin $\nu={1\over3}$ state of unpaired electrons, 
the increase of $2l$ by unity from the value corresponding to a 
full CF-LL$_0$ creates not one but two QH's, as predicted by 
Eq.~({\ref{eqEffg}) for our picture of Laughlin-correlated pairs.
Note that the same is true for the finite-size Jain $\nu={2\over5}$ 
states with two CF LL filled; however, no combination $(N,2l)$ 
corresponds to a  single QH in a finite-size paired Laughlin state 
(the condition $g_0^*=N_2+1$ leads to a half integral value of $g$), 
while for the Jain $\nu={2\over5}$ state it occurs for even $N$, 
at $2l={1\over2}(5N-7)$.
Similarly as in Fig.~\ref{fig7}, the first excited band above the 
2QH states contains an additional QE--QH pair, and it extends to
$L=(3l_0^*-3)+l_1^*=N$, exactly as marked in frame (a).

At $2l=(2N-3)-1$ no states can have $E=0$, but the lowest band
is expected to contain two QE's in CF-LL$_1$ of degeneracy
$g_1^*=N_2$.
Indeed, in spectra (b) and (c) obtained for $N=14$ and 16, the 
low-energy bands at $L=N_2-2$, $N_2-4$, \dots\ can be found as 
expected (although they are not as well resolved as the QH bands).

What is the electric charge $\mathcal{Q}$ of the QE's and QH's?
Being proportional to the LL degeneracy, it can be obtained from 
the ratio of $g^*$ and $g=N/\nu$ calculated in the $N\rightarrow
\infty$ limit.
For a Jain $\nu=n/(2pn+1)$ state of $n$ completely filled CF-LL's, 
the degeneracy of each one is $g^*=N/n$, which leads to the well-known 
result $\mathcal{Q}/e=g^*/g=(2pn+1)^{-1}$.
For the present case, $g=2N$, $g^*=N_2={1\over2}N$, and the 
result is precisely what should be expected for a $\nu={1\over8}$
state of $2e$-charged boson pairs\cite{halperin83}
\begin{equation}
   \mathcal{Q}=e/4.
\end{equation}

\subsection{Spectra for odd particle numbers}

If Halperin's\cite{halperin83} picture could be simply extended 
to finite $\nu={1\over2}$ systems with odd electron numbers $N$, 
they would contain $N_2={1\over2}(N-1)$ pairs and $N_1=1$ 
unpaired electron, forming a two-component Laughlin-correlated 
fluid.\cite{fivehalf}
What actually happens is presented in two sample energy spectra 
in Fig.~\ref{fig9}(a,b).
\begin{figure}
\resizebox{3.4in}{1.50in}{\includegraphics{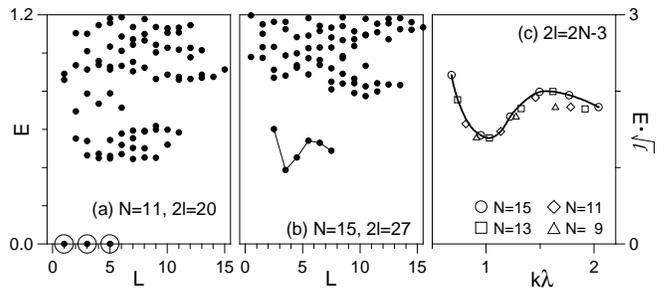}}
\caption{
\label{fig9}
   The same as Figs.~\ref{fig7} and \ref{fig8} but for odd 
   particle numbers $N$ and for $2l=2N-2$ (a) and $2l=2N-3$ (b).
   Energy dispersion (c) is for the pair-breaking band 
   marked in frame (b).
   $\lambda$ is the magnetic length.}
\end{figure}

At $2l=(2N-3)+1$ there is a band of $E=0$ states that indeed 
correspond to a pair of QH's of the two-component fluid.
In the CF picture, each QH has $l_0^*={1\over2}(N_2+1)$ which
gives the total $L=N_2$, $N_2-2$, \dots, exactly as obtained 
for $N=11$ in frame (a).

At $2l=2N-3$ no $E=0$ states occur, and the numerical results 
for different $N$ always show a band at $L={5\over2}$, ${7\over2}$, 
\dots, ${1\over2}N$, that seems to describe dispersion of an 
excitonic state of a pair of QP's of opposite charge.
This becomes more convincing in Fig.~\ref{fig9}(c), where the 
data obtained for different $N$ is plotted together as a function 
of wavevector $k$, and a clear magnetoroton-type minimum appears
at $k\approx1.0\,\lambda^{-1}$.
Remarkably, the values $l={1\over4}(N\pm5)$ of the QP angular 
momenta that would explain the observed range of $L$ do not 
agree with the prediction of a Laughlin-correlated state with
$N_2={1\over2}(N-1)$ and $N_1=1$.
Nevertheless, knowing their angular momenta is enough to predict
the charge $\pm e/4$ of these (not completely identified) QP's.

The reason why this low-energy band cannot be described by a 
two-component CF model (for any combination of $N_1$ and $N_2$, 
not just the one with $N_1=1$) is that they are not pair--pair 
or pair--electron, but pair-breaking excitations introduced by 
Greiter {\sl et al.}\cite{greiter91}
Such excitations generally occur in paired systems and they 
are expected to be charge-neutral (despite being fermions) which 
explains their continuous energy dispersion in a magnetic field.
Still, the above discussion suggests that it should be possible 
to decompose them into a pair of more elementary, charged QP's.

\section{Relevance to the $\nu={5\over2}$ FQH state}

Earlier diagonalization studies\cite{greiter91,rezayi00,morf98,%
fivehalf} using Coulomb pseudopotential in LL$_1$ showed the $L=0$ 
ground states with a gap at $2l=2N-3$ but no clear indication of 
QP excitations identified\cite{read96} in the spectra of the model 
three-body repulsion $W$.
As shown in the top frames of Fig.~\ref{fig10} obtained for $N=12$
electrons, the magnetoroton QE--QH band and of the two-QE bands 
can indeed hardly be found in these spectra due to mixing with 
higher states, and only the two-QH bands are well separated.
\begin{figure}
\resizebox{3.4in}{2.81in}{\includegraphics{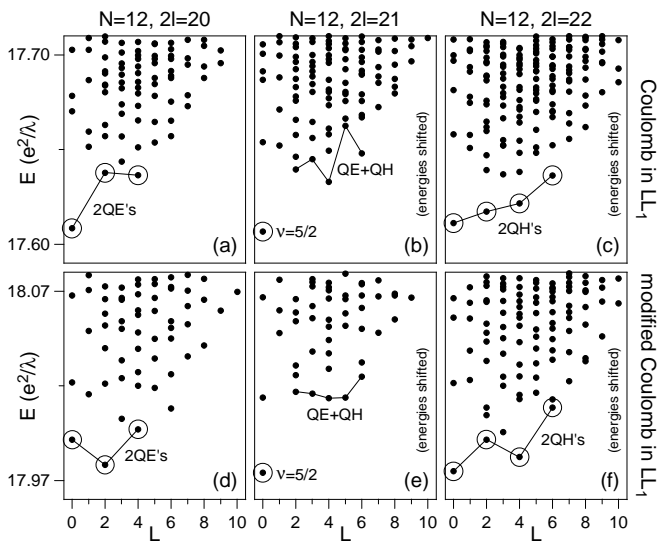}}
\caption{
\label{fig10}
   (a,b,c) 
   Energy spectra similar to those in Figs.~\ref{fig7}--\ref{fig9}, 
   but obtained for the Coulomb pair pseudopotential of the first 
   excited LL in Fig.~\ref{fig1}(b).
   (d,e,f) 
   Spectra of the same systems obtained for parameter $V_2(1)$ 
   increased by 9\%.}
\end{figure}

The problem with the identification of the Coulomb $\nu={1\over2}$
ground state in LL$_1$ with the Moore--Read (or any other) trial 
state is that the former is very sensitive to the relative values 
of the leading pseudopotential coefficients, while the exact form 
of $V_2(\mathcal{R}_2)$ depends (at least in principle) on the 
layer width $w$ in experiments, and on $N$ in finite-size 
calculations.
As to the width-dependence, it turns out that increasing $w$ 
from zero to realistic experimental values only weakly affects 
the nearly harmonic behavior of $V_2(\mathcal{R}_2)$ at short 
range that is essential for the avoidance of the $\mathcal{R}_3
=3$ triplet state.
As a result, the $\nu={5\over2}$ wavefunction in experimental 
systems depends much less on the width than, e.g., the excitation 
gap controlled by the magnitude of $V_2$.

On the other hand, the strong dependence of correlations on 
$\alpha\sim{1\over2}$ in finite systems is clear in 
Figs.~\ref{fig3}(a), \ref{fig5}(a), and \ref{fig6}, and it 
is in contrast with the behavior at $\alpha\sim0$ or 1, 
corresponding to the much less sensitive finite-size FQH 
states in LL$_0$ and CF-LL$_1$.
Remarkably, the gap above the incompressible ground state at 
$2l=2N-3$ persists\cite{qepair1} over a wide range of $\alpha$ 
despite even a large distortion of its wavefunction, while the 
QP excitations quickly mix with the continuum of higher states 
when $V_2$ becomes too sub- or superharmonic at short range.

A major problem with the calculations on a sphere is the 
size-dependence (\ref{eqLIM}) of the critical value of $\alpha$ 
at which the avoidance of $\mathcal{R}_3=3$ occurs.
It is clearly visible in the plots of squared overlaps $\zeta_u
(\alpha)=|\left<\phi_\alpha|\psi_u\right>|^2$ with the eigenstates 
$\phi_\alpha$ of $U_\alpha$, calculated for the corresponding 
eigenstates $\psi_u$ of various other interactions $u$: 
three-body repulsion $W$ and electron and QE pair pseudopotentials 
$V_2$ in LL$_0$, LL$_1$, and CF-LL$_1$, respectively.
For LL$_1$, the overlaps $\zeta_{\rm \,LL1}$ have been calculated 
for both narrow ($w=0$) and wide ($w=3.5\,\lambda$; e.g., $w=20$~nm 
at $B=20$~T) layers. 
Note also that the eigenstate or $W$ used in the calculation of 
overlaps is automatically properly symmetrized (in the original
form\cite{moore91} it is not\cite{rezayi00}).

In Fig.~\ref{fig11} we plot the overlaps for the lowest $L=0$
states at $2l=2N-3$.
\begin{figure}
\resizebox{3.4in}{1.61in}{\includegraphics{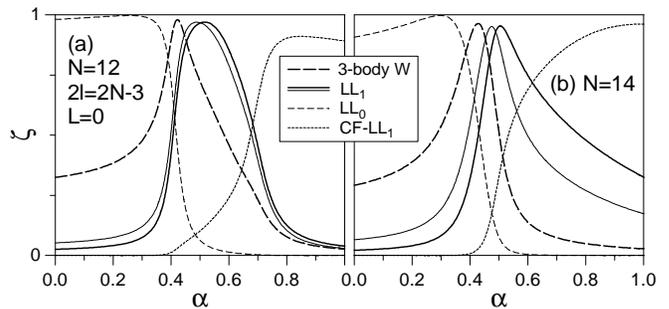}}
\caption{
\label{fig11}
   Squared overlaps $\zeta$ of the lowest $L=0$ eigenstate of 
   pair interaction $U_\alpha$ defined by Eq.~(\ref{eqU}) 
   calculated on a sphere at $2l=2N-3$, with the corresponding 
   eigenstates of three-body repulsion $W$ (Moore--Read state), 
   electron interaction in the lowest and excited LL (the narrow 
   solid line is for layer width $w=3.5\,\lambda$), and QE 
   interaction in the Laughlin $\nu={1\over3}$ state, plotted 
   as a function of $\alpha$.
   Frames (a) and (b) correspond to $N=12$ and 14 particles.}
\end{figure}
Clearly, the exact Moore--Read eigenstate of $W$ is an excellent
ground state of $U_\alpha$ at $\alpha\approx0.425$.
So is the ground state of Coulomb pair interaction in LL$_1$,
but at a different $\alpha\approx0.5$.
The disagreement between these two values of $\alpha$ does not
disappear in wide samples, as inclusion of $w$ even as large as
$3.5\,\lambda$ does not noticeably change the Coulomb $\nu={5\over2}$ 
ground state.
Specifically, the overlaps between the Moore--Read state and the
Coulomb $\nu={5\over2}$ ground state calculated for $N=14$ are 
only $|\left<\psi_W|\psi_{\rm \,C1}\right>|^2=0.48$, 0.58, and 
0.71 for $w/\lambda=0$, 1.75, and 3.5, respectively

The behavior $\zeta_{\rm QE}(\alpha)$ plotted with narrow dotted 
lines is also noteworthy.
The QE--QE interaction at a half-filling can be described by $U_1$ 
quite well for $N=14$ (where the calculations indicate a finite-size 
$L=0$ ground state with a gap) and somewhat worse for $N=10$ (where 
the ground state is compressible).
But even more interestingly, the Moore--Read state appears nearly 
orthogonal to the QE states (the exact value for $N=14$ is $|\left<
\psi_W|\psi_{\rm QE}\right>|^2=0.03$), which we interpret as yet 
another strong indication against the QE pairing at $\nu={3\over8}$.

In Fig.~\ref{fig12} we plot similar overlaps calculated for various
excitations.
\begin{figure}
\resizebox{3.4in}{2.53in}{\includegraphics{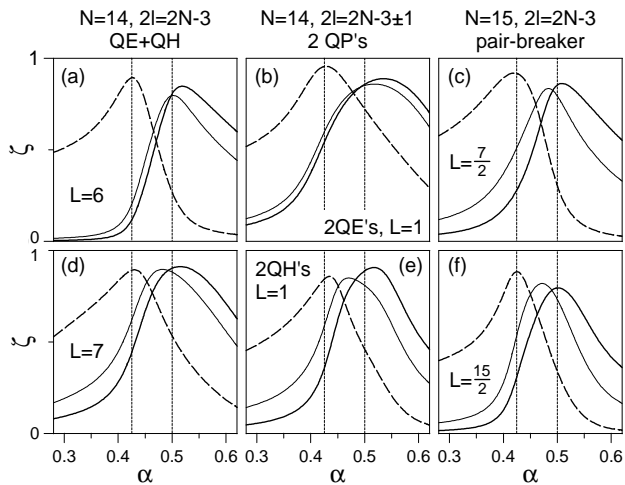}}
\caption{
\label{fig12}
   Similar to Fig.~\ref{fig11} (solid and dashed lines mean the
   same) but for different low-energy states at $2l=(2N-3)$ and 
   $(2N-3)\pm1$, corresponding to pair QE and QH states and the
   pair-breaking excitation of the three-body interaction $W$.}
\end{figure}
Frames (a,d) correspond to a QE--QH pair, (b) to two QE's, 
(e) to two QH's, and (c,f) to the pair-breaking neutral-fermion 
excitation.
We only show the curves for the QE--QH states at $L=6$ and 7 near 
the magnetoroton minimum, for two-QE and -QH states at small $L=1$
(corresponding to large QP--QP separation for which the curves 
are less dependent on QP--QP interaction effects), and for the 
pair-breaker at $L={7\over2}$ near the energy minimum and at 
a large $L={15\over2}$.
All frames show similar behavior to Fig.~\ref{fig11}, only the
disagreement between the eigenstates of $W$ and the Coulomb 
eigenstates is more pronounced.
The QP excitations of the three-body repulsion $W$ remarkably well 
describe actual excitations of a system with a two-body interaction 
$U_\alpha$.
However, not for the value of $\alpha$ corresponding to the Coulomb 
interaction in LL$_1$ (regardless of the layer width).
The overlaps between eigenstates of $W$ and the electron eigenstates 
in LL$_1$ are even lower than those for the Moore--Read state.
The specific values for $N=14$ and $w=0$ (and for $w=3.5\,\lambda$ 
in parentheses) are: $|\left<\psi_W|\psi_{\rm C1}\right>|^2=0.03$, 
0.00, 0.27, 0.19, 0.12, 0.46 (0.03, 0.02, 0.39, 0.31, 0.20, 0.60) 
for the $L=2$, 3, \dots, 7 states of the magnetoroton QE+QH band, 
0.47, 0.16, 0.07 (0.52, 0.28, 0.14) for the $L=1$, 3, 5 states
of two QE's, and 0.39, 0.12, 0.39, 0.27 (0.53, 0.17, 0.64, 0.32) 
for the $L=1$, 3, 5, 7 states of two QH's, respectively.
The values for the pair-breaking band for $N=13$ are: 0.45, 0.19, 
0.41, 0.31, 0.34 (0.56, 0.34, 0.44, 0.46, 0.47) for $L={5\over2}$, 
${7\over2}$, \dots, ${13\over2}$, respectively.
Such small overlaps preclude (indicated) interpretation of excited 
states in Fig.~\ref{fig10}(a) and (b) in terms of the QE's of $W$.

This invokes the question raised in the introduction of whether
the Moore--Read trial state and its QP excitations are only an
elegant idea, not realized in known even-denominator FQH states 
(at $\nu={5\over2}$ or ${3\over8}$).
Fortunately, the disagreement appears to be largely artificial.
The size-dependence (\ref{eqLIM}) of $\alpha_0$ can be traced 
to the size-dependence of the pair amplitudes $\mathcal{G}_2
(\mathcal{R}_2)$ of the triplet eigenstates at $\mathcal{R}_3=3$, 
5, 6, \dots, directly caused by the surface curvature.
It is therefore only due to this curvature that (in finite 
systems on a sphere) $\alpha_0<{1\over2}$ is different from 
the value $\alpha={1\over2}$ appropriate for the Coulomb 
pseudopotential in LL$_1$.
This appears to be consistent with larger overlaps calculated 
for the Moore--Read state in toroidal geometry.\cite{rezayi00} 

At $N\rightarrow\infty$, we expect that $\alpha_0\approx{1\over2}$ 
in coincidence with the behavior of $V_2(\mathcal{R}_2)$ in the 
same limit, and that the energy spectra of Coulomb $V_2$ and model 
$W$ interactions should become similar.
To improve the agreement at $N\le14$, for which we were able 
to calculate the spectra, $V_2(1)$ must be slightly enhanced 
in accordance with Eq.~(\ref{eqLIM}).
E.g., for $N=12$ the near vanishing of $\mathcal{G}_3(3)$ at
$2l=2N-3$ occurs when is $V_2(1)$ increased by 9\% from its 
Coulomb value, in good agreement with the result of Morf.
\cite{morf98}
The $N=12$ electron energy spectra calculated for this 
interaction with marked features associated with the QP's 
are shown in bottom frames of Fig.~\ref{fig10}.

The above discussion yields the following statements:
(i) 
Finite-size calculations on a sphere using Coulomb pair 
interaction do not correctly reproduce correlations of 
an infinite $\nu={5\over2}$ state. 
They use pseudopotentials corresponding to $\alpha\approx
{1\over2}$, different from $\alpha_0<{1\over2}$ leading to 
the avoidance of $\mathcal{R}_3=3$.
The $\alpha=\alpha_0={1\over2}$ coincidence is probably 
recovered for $N\rightarrow\infty$ which would mean that 
the real, infinite systems at $\nu={5\over2}$ do have the 
``$\mathcal{R}_3>3$'' correlations while the correlations 
in finite systems are different and size-dependent.
(ii)
In finite systems, correct ``$\mathcal{R}_3>3$'' correlations 
are recovered if the pair pseudopotential is appropriately 
enhanced at short range.
(iii)
Assuming that that the $\alpha=\alpha_0={1\over2}$ coincidence
is restored in infinite systems (or in different, e.g., toroidal 
geometry), the equivalence of Coulomb and $W$ interactions at 
half-filling is not limited strictly to the Moore--Read ground 
state.
The $(\pm e/4)$-charged QP's and the neutral-fermion pair-breaker 
identified in the spectra of $W$ accurately describe the low-energy 
charge excitations in the real (Coulomb) $\nu={5\over2}$ systems.
Although the effective interactions between QP's may lead to 
their binding or dressing (just as at $\nu={1\over3}$ QH's 
and ``reversed-spin'' QE's bind to form skyrmions), they are 
simple objects with an elegant interpretation in terms of 
Laughlin-like three-body correlations.

\section{Conclusion}

We have studied two- and three-body correlations in partially
filled degenerate shells for various interactions between the 
particles.
Variation of the relative strength of two leading pair 
pseudopotential coefficients drives the correlations through 
three distinct regimes.
The intermediate regime, corresponding to the nearly harmonic 
pseudopotential at short range, describes correlations among 
electrons in LL$_1$, particularly in the $\nu={5\over2}$ FQH 
state.

In contrast to the correlations between electrons in LL$_0$ or 
between Laughlin QE's in CF-LL$_1$ (whose pseudopotentials are 
strongly super- and subharmonic at short range, respectively),
the intermediate regime is not characterized by a simple 
avoidance of just one pair eigenstate corresponding to the 
strongest anharmonic repulsion.
Instead, we have shown that near the half-filling the low-energy 
states for such interactions have simple three-body correlations.
In resemblance of Laughlin pair correlations, they consist of
the maximum avoidance of the triplet state with the smallest 
relative angular momentum $\mathcal{R}_3=3$, i.e., with the 
smallest area spanned by the three particles (in analogy to pair 
correlations, avoidance means here the minimization of a 
triplet amplitude).

In particular, at exactly half-filling, this corresponds to 
the fact\cite{greiter91} that the Moore--Read ground state is 
the zero-energy eigenstate of a model short-range three-body 
repulsion $W$ with the only pseudopotential parameter at 
$\mathcal{R}_3=3$.
The Moore--Read ground state is a three-body analog of the 
Laughlin $\nu={1\over3}$ state with $\mathcal{R}_2>1$.
It is separated by a finite excitation gap from a magnetoroton 
band with a minimum at $k\approx1.5\,\lambda^{-1}$.
Its elementary excitations are the $(\pm e/4)$-charged QP's
(that naturally occur for the Halperin\cite{halperin83} state 
with Laughlin correlations between pairs) and the pair-breaking 
excitation.
The bands of few-QP states near half-filling are well 
described by a CF picture appropriate for Laughlin pair-pair 
correlations.

Finally, the problem of numerical calculations on a sphere 
associated with the surface curvature is addressed.
It is found that finite-size models using Coulomb interaction 
between electrons do not correctly reproduce correlations 
of the $\nu={5\over2}$ FQH state due to the distortion of 
triplet wavefunctions.
Especially for the excitations of the $\nu={5\over2}$ ground
state, the overlaps with the Moore--Read-like correlated 
states are rather small.
However, it is argued that the $\nu={5\over2}$ FQH state 
observed experimentally in narrow systems is described much 
better by the Moore--Read trial state than could be expected 
from the calculation of overlaps in small systems.
Consequently, the origin of its incompressibility is precisely
the avoidance of the $\mathcal{R}_3=3$ triplet state, and its 
elementary excitations are the $(\pm e/4)$-charged QP's
(although more complex excitations, such as bound QP states, 
might be the lower-energy charge carriers in realistic systems).

\acknowledgments

This work was supported in part by grant DE-FG 02-97ER45657 
of the Materials Science Program -- Basic Energy Sciences of 
the U.S. Dept.\ of Energy.  
AW acknowledges support by the Polish Ministry of Scientific 
Research and Information Technology under grant 2P03B02424.

\end{document}